\documentclass[american,aps,showpacs,twocolumn]{revtex4}
\usepackage[T1]{fontenc}
\usepackage[latin9]{inputenc}
\usepackage{babel}

\usepackage{amsmath}
\usepackage{graphicx}
\usepackage{amssymb}
\usepackage[unicode=true, 
 bookmarks=true,bookmarksnumbered=false,bookmarksopen=false,
 breaklinks=false,pdfborder={0 0 1},backref=false,colorlinks=false]
 {hyperref}
\hypersetup{pdftitle={A New Class of Spin Projection Operators},
 linkcolor=black, citecolor=black, urlcolor=black, pdfstartview=XYZ, plainpages=false, pdfpagelabels}

\makeatletter
\@ifundefined{textcolor}{}
{%
 \definecolor{BLACK}{gray}{0}
 \definecolor{WHITE}{gray}{1}
 \definecolor{RED}{rgb}{1,0,0}
 \definecolor{GREEN}{rgb}{0,1,0}
 \definecolor{BLUE}{rgb}{0,0,1}
 \definecolor{CYAN}{cmyk}{1,0,0,0}
 \definecolor{MAGENTA}{cmyk}{0,1,0,0}
 \definecolor{YELLOW}{cmyk}{0,0,1,0}
 }

\usepackage{amsgen}\usepackage{amsbsy}\usepackage{amsfonts}

\usepackage{babel}

\makeatother

\begin{document}

\title{A new class of spin projection operators for 3D models}

\author{Antonio Accioly}\email{accioly@cbpf.br}
\affiliation{Laborat\'{o}rio de F\'{\i}sica Experimental (LAFEX), Centro Brasileiro de Pesquisas F\'{i}sicas (CBPF), Rua Dr. Xavier Sigaud 150, Urca, 22290-180, Rio de Janeiro, RJ, Brazil}
\affiliation{Instituto de F\'{\i}sica Te\'{o}rica (IFT), S\~{a}o Paulo State University (UNESP), Rua Dr. Bento Teobaldo Ferraz 271, Bl. II - Barra Funda, 01140-070 S\~{a}o Paulo, SP, Brazil}
\author{Jos\'{e} Helay\"{e}l-Neto}\email{helayel@cbpf.br}
\affiliation{Laborat\'{o}rio de F\'{\i}sica Experimental (LAFEX), Centro Brasileiro de Pesquisas F\'{i}sicas (CBPF), Rua Dr. Xavier Sigaud 150, Urca, 22290-180, Rio de Janeiro, RJ, Brazil}
\author{Bruno Pereira-Dias}\email{bpdias@cbpf.br}

\affiliation{Laborat\'{o}rio de F\'{\i}sica Experimental (LAFEX), Centro Brasileiro de Pesquisas F\'{i}sicas (CBPF), Rua Dr. Xavier Sigaud 150, Urca, 22290-180 Rio de Janeiro, RJ, Brazil}

\author{Carlos Hernaski}\email{hernaski@unifap.br}
\affiliation{Universidade Federal do Amapá (Unifap), Rod. Juscelino Kubitschek
KM-02, Jardim Marco Zero, 68.902-280 Macapá, AP, Brazil}














\pacs{11.10.Kk, 04.60.Kz, 04.60.Rt, 04.50.Kd, 04.90.+e}
\begin{abstract}
A new set of projection operators for three-dimensional models are
constructed. Using these operators, an uncomplicated and easily handling
algorithm for analysing the unitarity of the aforementioned systems
is built up. Interestingly enough, this method converts the task of
probing the unitarity of a given 3D system, that is in general a time-consuming
work, into a straightforward algebraic exercise; besides, it also
greatly clarifies the physical interpretation of the propagating modes.
In order to test the efficacy and quickness of the algorithm at hand, the unitarity
of some important and timely higher-order electromagnetic (gravitational)
systems augmented by both Chern-Simons  and higher order Chern-Simons terms are investigated. 
\end{abstract}
\maketitle

\section{Introduction}

The well-known complexities of 4D field theory have often forced theorists
to test models in lower-dimensional spaces. In general, the foundations
of such models have been obtained only by projection from the physical
dimension which, of course, cannot shed light on the subtleties inherent
to a particular dimension. In this vein --- as it was pointed out by
Binegar \cite{1} with good reason --- an independent development
of the theories in their native dimension is required since a theorist
is not supposed to be omniscient. 

In this sense, three-dimensional
theories deserve a special attention due to its closeness to reality.
Fortunately, planar physics has undergone a remarkable development
in the last few decades. A host of new experimental results coming
mainly from condensed matter physics and the accompanying rapid convergence
of theoretical ideas have brought to the subject a new coherence and
have also raised new interests. Among the so many and interesting
planar models that have been investigated, it is worth mentioning
the graphene \cite{2}. This genuinely planar carbon system seems likely to be a good framework
for the verification of ideas and methods developed in quantum (gauge-)
field theories. Consequently, we hopefully expect that the techniques
of QED$_{3}$ when applied to this low-dimensional condensed-matter
model lead to new and relevant results \cite{3,4}. As far as gravity
is concerned, the reason for doing research on planar gravity is quite
amazing: (2+1)-dimensional gravity has a direct physical relevance
to modeling phenomena that are actually confined to lower dimensionality.
In fact, gravitational physics in the presence of straight cosmic
strings (infinitely long, perpendicular to a plane) is adequately described
by three-dimensional gravity \cite{5}. We remark that the causality
puzzles raised by `Gott time machines' were solved with the help of
this lower dimensional model \cite{6}.

On the other hand, three-dimensional electromagnetic (gravitational)
models enlarged by a Chern-Simons term have been the object of much
attention. In the vector case this term gives mass to the photon in
a gauge-invariant way; while, for planar gravity the Chern-Simons
term is responsible for the presence of a propagating parity-breaking
massive spin-2 mode in the spectrum of the model \cite{7}.

 Recently, both higher-order electromagnetic and gravitational models have enjoyed
a revival of interest. Indeed, the fourth-dimensional theory of quantum
electrodynamics proposed by Lee and Wick (LW) with the purpose of
understanding the finite electromagnetic mass splitting of mesons,
prior to QCD was established \cite{8,9}, has been rather explored
as a kind of toy model for the more complex dynamics of the LW Standard
Model, i.e., the model in which Grinstein, O'Connel and Wise, building
on the pioneering work of  LW, introduced non-Abelian LW gauge
theories \cite{10,11,12,13,14,15,16,17,18,19}; whereas, just about
three years ago, Bergshoeff, Hohm, and Townsend (BHT) \cite{20,21,22,23,24,25,26,27,28}
proposed a particular higher-derivative extension of the Einstein-Hilbert
action in three spacetime dimensions whose linearized version is a rare example of a fourth-order system that is not pestered by ghosts \cite{29}. Besides, a canonical analysis of the quadratic curvature part of the BHT system done by Deser \cite{29} establishes its  
weak field limit as both ghost-free and power-counting UV finite,
thus violating standard folklore in the extreme.

The preceding considerations naturally suggest that investigations
into general 3D higher-order electromagnetic (gravitational) models
with a Chern-Simons term, are welcome. The introduction of higher-derivatives,
nevertheless, could in principle jeopardize the unitarity of the models.
It would thus be very convenient, in the spirit of paragraph one,
to devise an easily handling procedure, specific to planar models,
which allowed, on physical grounds, a constructive and meaningful
discussion of the unitarity of generic 3D electromagnetic (gravitational)
models, in an uncomplicated way.

In this paper, the aforementioned procedure is constructed by means
of a new basis of spin operators, specific to 3D models, which allows
a Lagrangian decomposition into spin components.

The ideas underlying our theoretical framework are described in Section
\ref{sec:Algorithm}. We start off by building up a new class of spin projectors for
3D models and then discuss how to obtain the propagator for these
models via the mentioned operators. The procedure for probing the
unitarity of the 3D models is constructed afterwards. Two important
and timely higher-derivative systems enlarged by both Chern-Simons and higher derivative Chern-Simons terms
are employed in Section \ref{sec:Examples} to illustrate the level of generality and quickness of the method. 
In Section IV it is shown that the expressions ``closure'' and ``completeness'' cannot be used interchangeably, as far as  projection operators are concerned. In other words, the use of a closed set that does not obey the axiom of completeness leads undoubtedly to false physical results. Consequently,  the fact that the set constituted by our spin operators is  indeed complete  guarantees  that the physical results obtained by means of them are  reliable.  We also discuss in this section whether  the addition of a  Chern-Simons term to a  nonunitary  
higher-derivative model can convert it into a unitary system.  Some of the more technical results are
gathered in the  Appendix \ref{sec:ProjectionOperatorRelationAppendix}.

In our conventions the Greek letters denote spacetime indices, the
metric signature is (+1, -1, -1), $\epsilon_{012}=+1$, where $\epsilon_{\mu\nu\rho}$
is the Levi-Civita symbol, and $\hbar=c=1$

\section{Prescription for probing the unitarity of 3D models}\label{sec:Algorithm}

In the analysis of quantum aspects of any field theory, considerable
interest is devoted to the description of the particle spectrum and
the relativistic quantum properties of scattering processes of
the theory under investigation. Some of these issues may be understood
by means of the analysis of the propagator of the theory, which is
obtained by the inversion of the wave operator. Accordingly, it is
of great and fundamental importance to perform this inversion judiciously.
We shall begin by looking for a suitable basis for the linear operators
acting on the fields of the model. Using this basis, a generic expression
for the propagator will then be constructed. Finally, a procedure
for analyzing the unitarity of the 3D models, based on the preceding
ingredients, will be worked out. 

\subsection{A new set of spin projection operators for 3D models}

\label{sec:A-new-set}

We start off by searching for a basis for the vectorial space of the
wave operators. The vector space where these operators act is formed
by finite-dimensional representations of the Lorentz group. In 4D,
for instance, it is always possible to decompose these vector spaces
in a direct sum of subspaces with well defined spin since locally
the Lorentz group can be regarded as the tensor product $SU\left(2\right)\otimes SU\left(2\right)$.
Besides, the only mapping operators that can be built among these
projectors are those associated with the same spin. In fact, the existence
of mapping operators implies in a bijection between spaces which can
be achieved if and only if they have the same dimension. However,
the construction of operators that map a spin space into a subspace
with a different spin can only be realized by decomposing the larger
spin-space into a direct sum of subspaces defined by preferential
vectors. The explicit construction of the spin projectors for fields
of arbitrary rank can be made by appealing to the tensor product of
projectors of vector fields according to the rules of group representations.
The spin projectors, in 4D, that decompose the vector fields can be
explicitly constructed using the Minkowski metric and partial derivatives,
as follows

\begin{subequations}\begin{align}
P\left(0\right)_{\mu\nu} & \equiv\omega_{\mu\nu}=\frac{\partial_{\mu}\partial_{\nu}}{\Box},\quad\Box=\partial_{\mu}\partial^{\mu},\\
P\left(1\right)_{\mu\nu} & \equiv\theta_{\mu\nu}=\eta_{\mu\nu}-\omega_{\mu\nu}.\end{align}
\end{subequations} A careful analysis of the preceding equations
allows to conclude that the spin projectors and mapping operators
of spin subspaces with the same dimension should also be built solely
with the metric and partial derivatives, which leads us to the well
known Barnes-Rivers operators \cite{30,31,32}. It is worth noticing
that if extra vectors are used in the construction of the models,
such as a background vector in Lorentz violating models, operators
with well defined spin will be insufficient to form a basis for the
wave operators \cite{33}.

On the other hand, the issue of the attainment of the wave operator
and, subsequently,  that of the propagator, for 3D models, need to be dealt
carefully. Why is this so? Because now we have both parity-conserving
(PC) and parity-violating (PV) models \cite{34}. Since the PC systems are defined by Lagrangians involving
only the metric and partial derivatives, the appropriate basis for
expanding the wave operator is, of course, that made up of the usual
3D Barnes-Rivers operators. The `mark' of the PV models,
i.e, the characteristic feature that enables us to recognize them,
is, in turn, the presence of the Levi-Civita tensor, which allows
us to define another vector linear operator,

\begin{equation}
S^{\mu\nu}=\epsilon^{\mu\nu\rho}\partial_{\rho}.\label{eq:csoperator}\end{equation}

\noindent Using this operator, we can enlarge the usual operator basis,
$\left\{ \theta,\omega\right\} $, in order to obtain a complete set
of linear operators $\left\{ \theta,\omega,S\right\}$ \cite{35}. It is worth noticing, however,
that $\theta$ is no longer a spin operator since in the massive case
the 3D spin corresponds to unitary representations of $SO\left(2\right)$,
that are one-dimensional. In fact, the operators $\theta$ and $\omega$
divide the three-dimensional space into a direct sum of two subspaces
with dimensions 2 and 1, in this order, which implies that $\theta$
does not project into a spin subspace. Consequently, it is impossible
to put a transparent and accurate physical interpretation on the excitation
modes related to the PV models if they are expressed in terms of the
basis $\left\{ \theta,\omega,S\right\} $. A successful way of dealing
with this problem would be to opt for a basis associated with the spin
of the particles in 3D. We discuss in the sequel how this basis can
be constructed.

We begin by recalling that in quantum field theory particles are identified
as unitary irreducible representations (irreps) of the Poincaré group.
This identification provides two quantum numbers for particles: mass
and spin. The spin is characterised by the unitary representations
of the little group of the representative momentum of the particle,
namely, the subgroup of the Lorentz group that leaves the representative
momentum unchanged.

In 4D, the task of identifying the spin operators is easier than in
3D. The reason why this is so comes from the fact that in 4D the spin of the massive
particles is given by unitary irreps of the group $SO\left(3\right)$.
Such representations are associated with the representations of the
group $SU\left(2\right)$, which is the covering group of $SO\left(3\right)$.
Furthermore, the fundamental representation of $SU\left(2\right)$
is equivalent to its complex conjugate. Therefore all the representations
of $SU\left(2\right)$ are real and, consequently, representations
of $SO\left(3\right)$ are univocally related to the $SU\left(2\right)$
representations. This implies that if the wave operator is decomposed
into operators that projects into well defined irreps of $SO\left(3\right)$,
they are automatically identified as operators with well defined spin.

In 3D, on the other hand, we have a distinct situation, since the
unitary representations of $SO\left(2\right)$ are associated with
$U\left(1\right)$. The fundamental representation of $U\left(1\right)$
is not equivalent to its complex conjugate. Since all representations
of $SO\left(2\right)$ are real they are not directly related to representations
of $U\left(1\right)$, but rather they should be identified with the
direct sum of the fundamental representation and its complex conjugate.
However, all the irreps of $SO\left(2\right)$ are two-dimensional
and can be associated with representations of $U\left(1\right)$ by
the complexification of the fields. Consider, for instance, the vectorial
representation of $SO\left(2\right)$. A transformation of $SO\left(2\right)$
 acting on a vector $A=\left(A_{1},A_{2}\right)$, yields 

\begin{equation}
\left(\begin{array}{cc}
\mbox{cos}\theta & -\mbox{sin}\theta\\
\mbox{sin}\theta & \mbox{cos}\theta\end{array}\right)\left(\begin{array}{c}
A_{1}\\
A_{2}\end{array}\right).\label{eq:rotacao}\end{equation}
 The normalized eigenvectors of this transformation are

\begin{eqnarray}
\lambda_{1} & = & \frac{1}{\sqrt{2}}\left(\begin{array}{c}
1\\
i\end{array}\right),\quad\lambda_{2}=\frac{1}{\sqrt{2}}\left(\begin{array}{c}
1\\
-i\end{array}\right).\end{eqnarray}
 So, we can define a basis for the 3D Minkowski space, with the characteristic
that each of its vectors spans one and only one 1D subspace that is
an eigenspace of the $U\left(1\right)$ transformations, i.e.,

\noindent \begin{align}
e\left(0\right)_{\mu} & =\left(\begin{array}{c}
1\\
0\\
0\end{array}\right)\\
e\left(+1\right)_{\mu}\equiv & \left(e_{1}\right)_{\mu}=\frac{1}{\sqrt{2}}\left(\begin{array}{c}
0\\
1\\
i\end{array}\right),\\
e\left(-1\right)_{\mu}\equiv & \left(e_{2}\right)_{\mu}=\frac{1}{\sqrt{2}}\left(\begin{array}{c}
0\\
1\\
-i\end{array}\right),\end{align}
where $e\left(0\right)$ is time-like, whereas $e_{1}$ and $e_{2}$
are space-like vectors.

\noindent Thence, under a suitable unitary change of variables, a
real vector field transforms like\begin{equation}
\left(\begin{array}{c}
A_{1}\\
A_{2}\end{array}\right)\longrightarrow\left(\begin{array}{c}
\bar{A}_{1}\\
\bar{A}_{2}\end{array}\right)=\frac{1}{\sqrt{2}}\left(\begin{array}{c}
A_{1}+iA_{2}\\
A_{1}-iA_{2}\end{array}\right).\end{equation}
 In this way the transformation of the vector $\bar{A}$ under the rotation
\eqref{eq:rotacao} is given by

\begin{equation}
\left(\begin{array}{cc}
e^{i\theta} & 0\\
0 & e^{-i\theta}\end{array}\right)\left(\begin{array}{c}
\bar{A}_{1}\\
\bar{A}_{2}\end{array}\right).\end{equation}

We conclude, therefore, that it is possible to make the identification
of the vectorial representation of $SO\left(2\right)$, $\triangle$,
as the direct sum of the $U\left(1\right)$ fundamental representation,
$\square$, and its complex conjugation, $\square^{*}$, i.e.,

\begin{equation}
\triangle\sim\square\oplus\square^{*}.\label{eq:rep-rel-so2-u1}\end{equation}

\noindent The $\theta$-operator is the identity in the space of the
representation $\triangle$. We may split this space in the direct
sum of one-dimensional subspaces. Keeping in mind the 3D spin representations,
spin projection operators may be associated with the basic complex vectors
$e_{1}$ and $e_{2}$, as follows \begin{eqnarray}
\rho^{\mu\nu} & = & -e_{1}^{\mu}\left(e_{1}^{\nu}\right)^{*},\label{eq:rho}\\
\sigma^{\mu\nu} & = & -e_{2}^{\mu}\left(e_{2}^{\nu}\right)^{*}.\label{eq:sigma}\end{eqnarray}

\noindent Here, $\rho$ is the projection operator associated with
$\square$, while $\sigma$ is related to $\square^{*}$. Note that
$\rho$ and $\sigma$ are related by complex conjugation, $\rho^{*}=\sigma$;
in addition, they are Hermitian and non-symmetric: \begin{equation}
\rho_{\mu\nu}=\left(e_{1}\right)_{\mu}\left(e_{1}\right)_{\nu}^{*}=\left(e_{2}\right)_{\nu}\left(e_{2}\right)_{\mu}^{*}=\sigma_{\nu\mu}.\label{eq:symmetryRho}\end{equation}

\noindent This completes the identification of the spin modes for
vector fields. It is also important to express the Chern-Simons operator
\eqref{eq:csoperator} in terms of the spin projection operators.
In order to do this, we note that $\varepsilon^{\mu\nu\rho}e_{\mu}^{1}e_{\nu}^{2}\frac{k_{\rho}}{\sqrt{k^{2}}}=-i$,
since $e_{1}$, $e_{2}$ and $\frac{k_{\rho}}{\sqrt{k^{2}}}$ are
normalized, which allows us to write

\begin{equation}
S_{\mu\nu}=-\sqrt{k^{2}}\left(\rho_{\mu\nu}-\sigma_{\mu\nu}\right).\end{equation}

One could wonder about the possibility of building mapping operators
between the subspaces defined by $k$ and $e_{1}$ or $k$ and $e_{2}$.
In reality, these mapping operators are unnecessary since they would
explicitly depend on $e_{1}$ and $e_{2}$. Actually, in the case
of Lorentz preserving models, the wave operator is constructed using
solely $\eta$'s, $\partial$'s and $\epsilon$'s.

Before going on, it is important to discuss the meaning of parity
as far as the 3D operators we have just analyzed are concerned. In
3D, the representation of the parity operator in the Minkowski vector
space is given by

\begin{equation}
P=\left(\begin{array}{ccc}
1 & 0 & 0\\
0 & 1 & 0\\
0 & 0 & -1\end{array}\right).\end{equation}
 As a result, we get from (\ref{eq:rho}) and (\ref{eq:sigma}) that
$P\rho P^{-1}=\sigma$ and $P\sigma P^{-1}=\rho$, which clearly shows
that, unlike what occurs in 4D, we cannot assign a definite parity
for a given spin. We remark that 
\begin{eqnarray}
P\omega P^{-1}=\omega, \; P\theta P^{-1}=\theta.
\end{eqnarray}
 \noindent In the special case of the PC models, the aforementioned relations
allow us to conclude that the massive excitation modes for nontrivial
spins are nothing but spin doublets.

After this little digression, let us build up the spin projection
operator for rank-2 tensors. For these tensors, we have

\begin{align}
(\underbar{1}\oplus & -\underbar{1}\oplus\underbar{0})\otimes(\underbar{1}\oplus-\underbar{1}\oplus\underbar{0})= \nonumber\\
 & \left(3\times\underline{0}\oplus2\times\underline{1}\oplus2\times-\underline{1}\oplus\underline{2}\oplus-\underline{2}\right),\end{align}

\noindent where the underlined numbers denote the spin, and the remaining
ones are related to the spin-multiplicity.

Now, a general rank-2 tensor, $T_{\mu\nu}$, may be written as product
of two vectors, say $A_{\mu}$ and $B_{\nu}$. Therefore, \begin{equation}
T_{\mu\nu}=A_{\mu}B_{\nu}.\end{equation}

\noindent Since a generic vector can be split in its spin components
$A_{\mu}\supset\left(1\oplus-1\oplus0\right)$, by means of the spin
projection operators, $\rho$, $\sigma$ and $\omega$, namely,

\begin{equation}
A_{\mu}=\left(\rho_{\mu\rho}+\sigma_{\mu\rho}+\omega_{\mu\rho}\right)A^{\rho},\end{equation}
 a generic rank-two tensor may also be decomposed in its spin components
as follows \begin{align}
T_{\mu\nu} & =(\rho_{\mu\rho}\rho_{\nu\sigma}+\rho_{\mu\rho}\sigma_{\nu\sigma}+\rho_{\mu\rho}\omega_{\nu\sigma}+\sigma_{\mu\rho}\rho_{\nu\sigma}+\sigma_{\mu\rho}\sigma_{\nu\sigma} \nonumber\\
 & +\sigma_{\mu\rho}\omega_{\nu\sigma}+\omega_{\mu\rho}\rho_{\nu\sigma}+\omega_{\mu\rho}\sigma_{\nu\sigma}+\omega_{\mu\rho}\omega_{\nu\sigma})T^{\rho\sigma}. \end{align}

\noindent Note that $\rho$, $\sigma$, and $\omega$ are associated
with spin +1, -1, and 0, respectively, which implies that $\rho\rho$,
$\rho\omega$, $\omega\rho$, $\rho\sigma$, $\sigma\rho$, $\omega\omega$,
$\sigma\omega$, $\omega\sigma$, $\sigma\sigma$ (with the indices
omitted) are associated with spin +2, +1, +1, 0, 0, 0, -1, -1, and
-2, in this order.

In the case of the graviton field $h_{\mu\nu}$, which is a symmetric
rank-2 tensor, the symmetrization of the operators above yields the
following spin projection operators \begin{subequations}\begin{align}
 & P^{hh}\left(+2\right)_{\mu\nu;\rho\sigma}=\rho_{\mu\rho}\rho_{\nu\sigma},\label{eq:P(+2)}\\
 & P^{hh}\left(-2\right)_{\mu\nu;\rho\sigma}=\sigma_{\mu\rho}\sigma_{\nu\sigma},\label{eq:P(-2)}\\
 & P^{hh}\left(+1\right)_{\mu\nu;\rho\sigma}=\frac{1}{2}\left(\rho_{\mu\rho}\omega_{\nu\sigma}+\rho_{\nu\rho}\omega_{\mu\sigma}+\rho_{\mu\sigma}\omega_{\nu\rho}+\rho_{\nu\sigma}\omega_{\mu\rho}\right),\\
 & P^{hh}\left(-1\right)_{\mu\nu;\rho\sigma}=\frac{1}{2}\left(\sigma_{\mu\rho}\omega_{\nu\sigma}+\sigma_{\nu\rho}\omega_{\mu\sigma}+\sigma_{\mu\sigma}\omega_{\nu\rho}+\sigma_{\nu\sigma}\omega_{\mu\rho}\right),\label{eq:P(-1)}\\
 & P_{11}^{hh}\left(0\right)_{\mu\nu;\rho\sigma}=\omega_{\mu\rho}\omega_{\nu\sigma},\\
 & P_{22}^{hh}\left(0\right)_{\mu\nu;\rho\sigma}=\frac{1}{2}\left(\rho_{\mu\rho}\sigma_{\nu\sigma}+\rho_{\nu\rho}\sigma_{\mu\sigma}+\rho_{\mu\sigma}\sigma_{\nu\rho}+\rho_{\nu\sigma}\sigma_{\mu\rho}\right).\end{align}
\end{subequations}

\noindent The preceding operators, of course, are Hermitian. In addition,
the projection operators associated with non-trivial spins are complex,
whereas those related to spin-$0$ are real because non-trivial spins
are non-trivial representations of $U\left(1\right)$, that are complex.
For a real Lagrangian, the complex structures (\ref{eq:P(+2)})-(\ref{eq:P(-1)})
alone cannot appear in the wave operator decomposition in terms of
the spin projection operators (this decomposition will be clarified
later). We can ensure, however, due to the Lorentz invariance of the
model, that projectors of the irreps of SO(2) will be present in the
wave operator. Such operators, usually known as Barnes-Rivers operators, are written in terms of $\theta$ and $\omega$. Using
the identity, $\theta=\rho+\sigma$, we may split the real Barnes-Rivers
operators into spin projection operators. Since $\rho^{*}=\sigma$,
the wave operator is obviously real.

Consider, for instance, the projector associated with a symmetric
and traceless rank-2 tensor. This operator projects into a non-trivial
and irrep of SO(2) and, therefore, it is two-dimensional, and can
be written as

\begin{equation}
P^{hh}\left(2\right)_{\mu\nu;\rho\sigma}=\frac{1}{2}\left(\theta_{\mu\rho}\theta_{\nu\sigma}+\theta_{\mu\sigma}\theta_{\nu\rho}\right)-\frac{1}{2}\theta_{\mu\nu}\theta_{\rho\sigma}.\label{spin projector}\end{equation}
 Now, taking into account that $\theta=\rho+\sigma,$ we obtain two
projectors in terms of $\rho$ and $\sigma$, one for each degree
of freedom of spin, i.e., \begin{equation}
P^{hh}\left(2\right)_{\mu\nu;\rho\sigma}=\rho_{\mu\rho}\rho_{\nu\sigma}+\sigma_{\mu\rho}\sigma_{\nu\sigma},\end{equation}
 clearly showing that the Barnes-Rivers operator $P^{hh}\left(2\right)$,
is nothing but a sum of spin +2 and spin -2 operators.
This process of decomposition can be repeated for all operators needed
to exhaust all the possibilities of contraction of the fields present
in the free Lagrangian. With this decomposition, the gravitational
Chern-Simons operator \begin{equation}
S_{\mu\nu;\rho\sigma}=\theta_{\mu\rho}S_{\nu\sigma}+\theta_{\mu\sigma}S_{\nu\rho}+\theta_{\nu\rho}S_{\mu\sigma}+\theta_{\nu\sigma}S_{\mu\rho},\end{equation}
  can be expressed
as\begin{equation}
S_{\mu\nu;\rho\sigma}=-4i\sqrt{k^{2}}\left(P\left(+2\right)_{\mu\nu;\rho\sigma}-P\left(-2\right)_{\mu\nu;\rho\sigma}\right).\end{equation}
 The other relations among the operators are listed in the Appendix \ref{sec:ProjectionOperatorRelationAppendix}.

\subsection{The propagator}

\label{subsec:The-propagator}

We are now ready to find the propagator and present afterwards the
algorithm for probing the unitarity of 3D electromagnetic (gravitational)
models. Consider, in this vein, a 3D Lagrangian $\mathcal{L}$ which
is a function either of a vector field $A_{a}$ or of a symmetric
rank-2 field, $h_{ab}$. In order to compute the propagator for the
model, we need beforehand the quadratic part of $\mathcal{L}$, i.e.,

\begin{equation}
\left(\mathcal{L}\right)_{2}=\frac{1}{2}\sum_{\alpha,\beta}\varphi_{\alpha}\mathcal{O}_{\alpha\beta}\varphi_{\beta},\label{eq:lagrangianaQuadratica}\end{equation}
where $\alpha,\beta$  represent vectorial or tensorial indices,
$\mathcal{O}_{\alpha\beta}$ is a local differential operator (the
wave operator) and $\varphi_{\alpha}$ encompasses the fundamental
quantum fields of the model. For gravity models, for instance,
this is accomplished by means of the weak field approximation of the metric, i.e.,
$g_{\mu\nu}=\eta_{\mu\nu}+h_{\mu\nu}$.

Using the identities of Appendix \ref{sec:ProjectionOperatorRelationAppendix},
we then expand the wave operator in the basis of the spin operators,
namely,

\begin{equation}
\mathcal{O}_{\alpha\beta}={\displaystyle \sum_{ij,J}a\left(J\right)_{ij}P_{ij}^{\varphi\varphi}\left(J\right)_{\alpha\beta}.}\label{eq:quadratic lagrangian}\end{equation}
 Here, $a\left(J\right)_{ij}$ are the coefficients of the expansion
of the wave operators. The diagonal operators, $P_{ii}^{\varphi\varphi}\left(J\right)$,
are operators that project the field $\varphi$ into its spin $J$.
Whereas the off-diagonal operators $(i\neq j)$ implement mappings
into the corresponding spin doublet subspace. The resulting spin
operators do obey the orthonormal multiplicative rules and the decomposition
of unity, i.e., \begin{align}
 & {\displaystyle \sum_{\beta}P_{ij}(I)_{\alpha\beta}P_{kl}(J)_{\beta\gamma}}=\delta_{jk}\delta^{IJ}P_{il}(I)_{\alpha\gamma},\\
 & {\displaystyle \sum_{i,J}P_{ii}(J)_{\alpha\beta}}=\delta_{\alpha\beta}.\end{align}
 This converts the task of inverting the wave operator (\ref{eq:quadratic lagrangian})
into an straightforward algebraic exercise. Indeed, all we have to
do is to invert the matrix of coefficients $a\left(J\right)_{ij}$.
Nevertheless, $a\left(J\right)_{ij}$ may be degenerate due to the
gauge symmetries of the model since the physical sources actually
may satisfy some constraints. These consistently appear in order to
inhibit the propagation of non-physical modes. The explicit expressions
for these constraints are given in terms of the left null-eigenvectors
$V_{j}^{(L,n)}$ of the degenerate coefficient matrices \cite{Sezgin:1979zf}

\noindent \begin{equation}
\sum_{\beta}V_{j}^{(L,n)}\left(J\right)P_{kj}\left(J\right)_{\alpha\beta}\mathcal{S}_{\beta}=0.\label{eq:constraintsSources}\end{equation}
 Nonetheless, since the propagator is saturated with the physical
sources, the correct procedure for the attainment of the propagator
is to invert any largest non-degenerate (for general values of momenta
$k$) sub-matrix of $a\left(J\right)_{ij}$. Accordingly, in order
to obtain the propagator, it suffices, in practice: (i) to delete rows
and columns of $a\left(J\right)_{ij}$ according to the number of gauge symmetries, which gives rise to  a matrix that we shall call $A\left(J\right)_{ij}$, and
  (ii) to invert $A\left(J\right)_{ij}$
and subsequently saturate this matrix with physical sources. As a result, the saturated propagator ($\Pi$) assumes the form \cite{Sezgin:1979zf}

\begin{equation}
\Pi=i\sum_{J,ij}\mathcal{S}_{\alpha}^{*}A(J)_{ij}^{-1}P_{ij}(J)_{\alpha\beta}\mathcal{S}_{\beta}'.\label{eq:propagatorDegenaraciesParityBreaking}\end{equation}
 It must be emphasized that since the physical sources satisfy the
constraints (\ref{eq:constraintsSources}), the propagator is gauge
independent. That is a great virtue of our method in comparison with
the methods that do not use orthonormal projection operators. Indeed,
in our procedure no gauge fixing is required.

\subsection{The prescription}

For the sake of simplicity, we shall divide our discussion about the
unitarity of the 3D models into two parts: one of them related to
the massive poles, the other concerning the massless ones. 
\begin{itemize}
\item \textit{massive poles} \\
 To ensure that there are neither ghosts nor tachyons in the propagation
mode of a given 3D model, we must require that at each simple pole
of the propagator $\left(k^{2}=m^{2}\right)$,

\begin{equation}
\Im m\mbox{Res}(\Pi|_{k^{2}=m^{2}})>0,\quad\mbox{and}\quad m^{2}\geq0.\label{eq:ghost/tachyon conditions for massive poles}\end{equation}

In the light of (\ref{eq:propagatorDegenaraciesParityBreaking}),
we come to the conclusion that the condition for the absence of ghosts
for each spin and for arbitrary sources is directly related to the
positivity of the matrices $\left(\sum A\left(J,m^{2}\right)_{ij}^{-1}P_{ij}\left(J\right)\right)_{\alpha\beta}$,
where $A\left(J,m^{2}\right)_{ij}^{-1}=\mbox{Res}\left.A\left(J,m^{2}\right)_{ij}^{-1}\right|_{k^{2}=m^{2}}$
is the matrix $A\left(J\right)_{ij}^{-1}$ with the pole extracted.
Furthermore, it can be shown that these matrices have only one non-vanishing
eigenvalue at the pole, which is equal to the trace of $\left.A^{-1}\left(J,m^{2}\right)\right|_{k^{2}=m^{2}}$.
Besides, the operators $P_{ij}\left(J\right)$ contribute only
with a sign $(-1)^{N}$, whenever calculated at the pole, where $N$
is the sum of the number of $\rho$'s and $\sigma$'s in each term
of the projector. This sign $\left(-1\right)^{N}$ can be understood
by noting that in rest frame of the particle, $\rho$ and $\sigma$
contribute with a minus sign, whereas $\omega$ contributes with a
positive sign. In summary, we may say that the conditions for absence
of ghosts and tachyons are such that for each massive single pole:
(i) $m^{2}>0$, and (ii) $(-1)^{N}\mbox{tr}A^{-1}\left(J,m^{2}\right)>0$.

\item \textit{massless poles}

The massless modes have some subtleties which requires an extra care.
Indeed, at first sight it seems that the basis of operators is not
well-defined for light-like momenta. However, the physical sources
constraints, that have its origin in the gauge symmetries of the model,
allow to find a well-defined expression for the saturated propagator,
even for light-like momenta. These physical constraints (\ref{eq:constraintsSources})
take the form $k_{\mu}\mathcal{S}^{\mu}=0$ for the electromagnetic
models and $k_{\mu}\mathcal{S}^{\mu\nu}=0$ for the gravitational
ones. Consequently, a convenient way of avoiding ghosts in the massless
modes is to rewrite the inverse of the wave operator in terms of the
following structures 
\begin{equation}
\omega_{\mu\nu}=\frac{k_{\mu}k_{\nu}}{k^{2}},\;\theta_{\mu\nu}=\eta_{\mu\nu}-\omega_{\mu\nu},\;\epsilon_{\mu\nu\rho},\; k_{\mu}.\label{eq:buildingblocksteta}
\end{equation}
 This task can be greatly facilitated by appealing to the relations
of the Appendix \ref{sec:ProjectionOperatorRelationAppendix}; in addition,
the sources must be expanded in a suitable momentum basis,

\begin{equation}
\mathcal{S}_{\mu}=c_{1}k_{\mu}+c_{2}q_{\mu}+c_{3}\epsilon_{\mu},\label{eq:sourceExpansionVector}\end{equation}
 \begin{eqnarray}
\mathcal{S}_{\mu\nu} &=&c_{1}k_{\mu}k_{\nu}+c_{2}\left(k_{\mu}\epsilon_{\nu}+k_{v}\epsilon_{\mu}\right)\nonumber\\
 &&+c_{3}\left(k_{\mu}q_{\nu}+k_{\nu}q_{\mu}\right)+c_{4}q_{\mu}q_{\nu}\nonumber \\
 &&+c_{5}\left(q_{\mu}\epsilon_{\nu}+q_{\nu}\epsilon_{\mu}\right)+c_{6}\epsilon_{\mu}\epsilon_{\nu}, 
\end{eqnarray}

where the $c_{i}$'s are complex coefficients, and\begin{subequations}\begin{align}
 & k_{\mu}=(k_{0},\vec{k}),\\
 & q_{\mu}=(k_{0},-\vec{k}),\end{align}
\end{subequations} with

\begin{subequations}\begin{align}
 & k^{2}=q^{2}=0,\label{eq:momentumBasis1}\\
 & k\cdot q=(k_{0})^{2}+(\vec{k})^{2},\\
 & k\cdot\epsilon=q\cdot\epsilon=0,\\
 & \epsilon^{2}=-1.\label{eq:momentumBasis4}\end{align}
\end{subequations}

The expansion (34)-(35) is the most general one for both vectors and symmetric rank-$2$ tensors
and must be supplemented by the sources constraints \eqref{eq:constraintsSources}.
Accordingly, the positivity of the residue of the propagator is assured
if \begin{equation}
\Im m\mbox{Res}(\Pi|_{k^{2}=0})\geq0.\end{equation}

\end{itemize}

\section{Testing the efficacy and quickness of the prescription}\label{sec:Examples}

In order to explicitly illustrate the generality and  simplicity of the proposed
method we analyze in the following  the unitarity of some higher-derivative electromagnetic (gravitational) models enlarged by both Chern-Simons and higher order Chern-Simons terms. We also comment, in passing, on some  interesting and remarkable properties of these systems.

\subsection{Higher-derivative electromagnetic models\label{sub:Maxwell-Podolski-Chern-Simons-Electrodynamics}}

We begin our study by checking the unitarity of the Lee-Wick-Chern-Simons model which is defined by the Lagrangian,

\begin{equation}
\mathcal{L}_\mathrm{LWCS}=-\frac{1}{4}F_{\mu\nu}F^{\mu\nu}-\frac{1}{4m^2}F_{\mu\nu}\square F^{\mu\nu}+\frac{\mu}{2}\epsilon^{\mu\nu\rho}A_{\mu}\partial_{\nu}A_{\rho},\label{MCS lagrangian}\end{equation}
 where $F_{\mu\nu}=\partial_{\mu}A_{\nu}-\partial_{\mu}A_{\nu}$, and $m$ ($\mu$) is a   parameter with dimension of mass.



Now, writing the  Lagrangian above in the form (26)  we promptly obtain the expression for the wave operator in momentum space, namely,  
\begin{equation}
\mathcal{O}_{\mu\nu}=\left(-k^{2}+\frac{k^{4}}{m^2}\right)\theta_{\mu\nu}-i\mu\epsilon_{\mu\nu\rho}k^{\rho}.\label{eq:waveOperator}\end{equation}

With the help  of the identities listed  in  the Appendix A,
  the wave operator \eqref{eq:waveOperator} may be expanded in  the 3D spin projection operators basis, as follows\begin{equation}
\mathcal{O}_{\mu\nu}={\displaystyle \sum_{ij,J}a\left(J\right)_{ij}P_{ij}^{AA}\left(J\right)_{\mu\nu},}\end{equation}
 where  \begin{equation}
a\left(0\right)=0,\end{equation}
 \begin{equation}
a\left(1\right)=\left(\begin{array}{cc}
-k^{2}+\frac{k^{4}}{m^2}+\mu\sqrt{k^{2}} & 0\\
0 & -k^{2}+\frac{k^{4}}{m^2}-\mu\sqrt{k^{2}}\end{array}\right).\end{equation}

It is worth noticing   that the spin-0 sector is completely degenerate, which is fully expected since
 the model
\eqref{MCS lagrangian} has a gauge symmetry \begin{equation}
A'_{\mu}=A_{\mu}+\delta A_{\mu}.\label{eq:gaugeSym}\end{equation}
 The term $\delta A_{\mu}$ can be easily  obtained  by noticing that it can be associated with
  the right null
eigenvalues of the matrices of the coefficients $V_{i}^{\left(R,n\right)}$ \cite{Sezgin:1979zf}, 

\begin{equation}
\delta\Psi_{\alpha}=\sum_{i,J,\beta}V_{i}^{\left(R,n\right)\psi}\left(J\right)P_{ij}^{\Psi\lambda}\left(J\right)_{\alpha\beta}f_{\beta}\left(J\right).\label{eq:transformacao de gauge}\end{equation}
 This result applies to every independent value of $j$ and $n$. For this example, 
we get
\begin{equation}
\delta A_{\mu}=\partial_{\mu}\left(\partial_{\nu}f^{\nu}\right),
\end{equation}
 where $f^{\nu}$ is an arbitrary function, as it should.

Interestingly,  the gauge symmetry of the model inhibits the propagation of the spinless mode; as a  consequence of this symmetry,
 there appears a source constraint that prevents  the propagation of this unphysical state.  It is trivial to see that  the general expression
\eqref{eq:constraintsSources}  reduces now to\begin{equation}
k_{\mu}\mathcal{S}^{\mu}=0,\label{eq:sourcConstEM}\end{equation}
 which is nothing but the familiar source conservation relation. 
 
 The inverse matrix of the spin-1 sector, on the other hand, reads

\begin{equation}
a\left(1\right)^{-1}=\frac{1}{\Delta}\left(\begin{array}{cc}
-k^{2}+\frac{k^{4}}{m^2}-\mu\sqrt{k^{2}} & 0\\
0 & -k^{2}+\frac{k^{4}}{m^2}+\mu\sqrt{k^{2}}\end{array}\right), \label{eq:spin-1-propagator-EM}\end{equation}
 where $\Delta=\left[\left(k^{2}-m^2\right)^{2}\frac{k^{2}}{m^4}-\mu^{2}\right]k^{2}=\left(\frac{k^{6}}{m^4}-2\frac{k^{4}}{m^2}+k^{2}-\mu^{2}\right)k^{2}$  is a quartic polynomial in $k^{2}$. As a result, it has four roots: one massless pole, and three massive ones which we shall call $m_{1}$, $m_{2}$, and $m_{3}$, respectively. Therefore, as far as the nature of  the roots are concerned, there exist precisely four distinct cases to be dealt with (see Fig. 1): (1) $\mu=0$ (Lee-Wick electrodynamics), (2) $0<\mu^{2}<\frac{4m^2}{27}$ (In this case there are three real positive masses.),  (3) $\mu^{2}>\frac{4m^2}{27}$ (Here there are  necessarily two complex roots, implying in the existence of tachyonic excitations.), and (4) $\mu^{2}=\frac{4m^2}{27}$ (In this case there appears a double pole; as a result, the model is non-unitary.). We discuss in the following  only the physical models, i.e., cases (1) and (2).

\begin{center}
\begin{figure}[h]

\begin{centering}
\includegraphics[width=7cm]{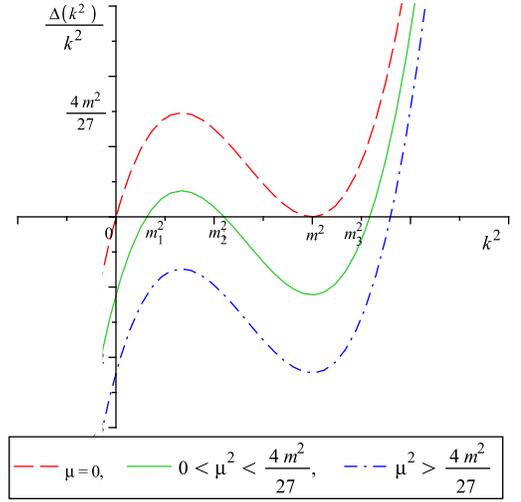} 
\par\end{centering}

\caption{Polynomial function $\frac{\Delta\left(k^{2}\right)}{k^{2}}$ versus $k^2$, where $\Delta\left(k^{2}\right)$
refers to  the denominator of the propagator \eqref{eq:spin-1-propagator-EM}. \label{fig:Pole-of-the}}

\end{figure}

\par\end{center}

\subsubsection{$\mu=0$ (Lee-Wick Electrodynamics)}

 The matrix of the coefficients is now given by 

\begin{equation}
a\left(1\right)=\left(\begin{array}{cc}
-k^{2}+\frac{k^{4}}{m^2} & 0\\
0 & -k^{2}+\frac{k^{4}}{m^2}\end{array}\right),\end{equation}
 while its   inverse  can be written as

\begin{equation}
a\left(1\right)^{-1}=\frac{1}{\left(k^{2}-m^2\right)k^{2}}\left(\begin{array}{cc}
1 & 0\\
0 & 1\end{array}\right). \end{equation}

Consequently, the absence of  tachyons and ghosts in the model is subordinated, respectively, to the following conditions \begin{equation}
m^{2}>0,\end{equation}
\begin{equation}
\left(-1\right)\mbox{tr}A(1,m^{2})^{-1}|_{k^{2}=m^{2}}=-1, \end{equation}
which clearly shows the presence of a non-tachyonic massive ghost in the system. 

For the massless pole, the  constraint $k_{\mu}\mathcal{S}^{\mu}=0$ allows us to write the saturated 
propagator  as

\textbf{\begin{equation}
\Pi=\frac{1}{\left(k^{2}-m^2\right)k^{2}}i\mathcal{S}^{*\mu}\mathcal{S}_{\mu}.\end{equation}
 }Expanding now the current $\mathcal{S}^{\mu}$ in the momentum
basis (\ref{eq:sourceExpansionVector}), yields

\begin{equation}
\Im m\mbox{Res}(\Pi|_{k^{2}=0})=\left|c_{1}\right|^{2}>0,\end{equation}
which allows us to conclude that the massless mode does not violate the unitarity. 

The wrong sign of Eq. (53) indicates an instability of the theory  at the classical level. From the quantum point of view it means that the theory in non-unitary. Luckily, these difficulties can be circumvented. Indeed, the classical instability can be removed by imposing a future boundary condition in order to prevent exponential growth of certain modes. However, this procedure leads to causality violations in the theory \cite{Coleman:1970}; fortunately, this acausality is suppressed below the scales associated with the   Lee-Wick particles. On the other hand, Lee and Wick argued that  despite the presence of the aforementioned degrees of freedom associated with  a non-positive definite norm on the Hilbert space, the theory could nonetheless be unitary as long as the new Lee-Wick particles obtain decay widths. There is no general proof of unitarity at arbitrary loop order for the Lee-Wick electrodynamics; nevertheless, there is no  known example of unitarity violation. Accordingly the  Lee-Wick electrodynamics is finite. Therefore, we need not be afraid of the massive spin-1 ghost. It is worth mentioning   that recently a  quantum bound on the Lee-Wick heavy particle mass was found that is of the order of the  mass of the $Z^0$ boson \cite{Accioly:2011zz}. Actually, this is  a very important result. Indeed, on the one hand, the knowledge of this parameter converts the aforementioned model into a predictable one; on the other, it introduces a natural scale for the model. As a result, we can estimate in the limit of static charge, for instance,  the distance in which the Lee-Wick potential departs from the usual Coulomb one. In a sense, this result  allows to ascertain  that only for small distances does the higher-derivative term  coming from the Lee-Wick model affect the well established results obtained within the context of the usual QED, as expected \cite{Accioly:2011zz}.

\subsubsection{$0<\mu^{2}<\frac{4m^2}{27}$}

Here $\Delta\left(k^{2}\right)=\left(k^{2}-m_{1}^{2}\right)\left(k^{2}-m_{2}^{2}\right)\left(k^{2}-m_{3}^{2}\right)k^{2}/m^4$, where $m_{1}$, $m_{2}$, and $m_{3}$ are the three real positive roots of $\Delta$. We assume without any loss of generality that $m_{1}<m_{2}<m_{3}$
(see Fig. \ref{fig:Pole-of-the}). On the other hand, the relations $\left(-1\right)\mbox{tr}A(1,m_{i}^{2})|_{k^{2}=m_{i}^{2}}>0$ 
 ($i=1,2,3,$) lead to the following inequalities 

\begin{eqnarray}
 & m_{1}: & \frac{\left(m_{1}^{2}-m^2\right)}{\left(m_{1}^{2}-m_{2}^{2}\right)\left(m_{1}^{2}-m_{3}^{2}\right)}<0,\label{eq:m1}\\
 & m_{2}: & \frac{\left(m_{2}^{2}-m^2\right)}{\left(m_{2}^{2}-m_{1}^{2}\right)\left(m_{2}^{2}-m_{3}^{2}\right)}<0,\\
 & m_{3}: & \frac{\left(m_{3}^{2}-m^2\right)}{\left(m_{3}^{2}-m_{1}^{2}\right)\left(m_{3}^{2}-m_{1}^{2}\right)}<0,\label{eq:m3}\end{eqnarray}
implying that $m_{1}^{2}<m^2$, $m_{2}^{2}>m^2,$ and $m_{3}^{2}<m^2$,
which, of course, contradicts the original assumption that $m_{1}<m_{2}<m_{3}$. Thence, we come to the conclusion that this model is plagued by ghosts.

\subsubsection{Lee-Wick-Chern-Simons  model enlarged by a higher derivative Chern-Simons extension}

Another interesting model can be built up from the Lee-Wick-Chern-Simons system  by adding to   the Lagrangian \eqref{MCS lagrangian}
the higher derivative Chern-Simons extension proposed by Deser and Jackiw \cite{Deser:1999} \begin{equation}
\mathcal{L}_\mathrm{ECS}=\frac{\lambda}{2}\epsilon^{\mu\nu\rho}\square A_{\mu}\partial_{\nu}A_{\rho}.\label{eq:HigherDerivativeCS}\end{equation}

Let us then check the unitarity of this curious model. Starting from the wave operator in momentum space,

 \begin{equation}
\mathcal{O}_{\mu\nu}=\left(-k^{2}+\frac{k^{4}}{m^2}\right)\theta_{\mu\nu}-i\left(\mu+\lambda k^{2}\right)\epsilon_{\mu\nu\rho}k^{\rho}, \label{eq:waveOpRC}\end{equation}

\noindent  it is straightforward to show that the spin-1 matrix of the
coefficients and its inverse are  respectively given by

\begin{widetext}

\begin{equation}
a\left(1\right)=\left(\begin{array}{cc}
-k^{2}+\frac{k^{4}}{m^2}+\left(\mu+\lambda k^{2}\right)\sqrt{k^{2}} & 0\\
0 & -k^{2}+\frac{k^{4}}{m^2}-\left(\mu+\lambda k^{2}\right)\sqrt{k^{2}}\end{array}\right), \end{equation}

\begin{equation}
a^{-1}\left(1\right)=\frac{1}{\Delta}\left(\begin{array}{cc}
-k^{2}+\frac{k^{4}}{m^2}-\left(\mu+\lambda k^{2}\right)\sqrt{k^{2}} & 0\\
0 & -k^{2}+\frac{k^{4}}{m^2}+\left(\mu+\lambda k^{2}\right)\sqrt{k^{2}}\end{array}\right),\label{eq:spin1EM}\end{equation}
\end{widetext} where \begin{equation}
\Delta=\left[\frac{k^{6}}{m^4}-\left(1+\lambda^{2}\right)\frac{k^{4}}{m^2}+\left(1-2\lambda\mu\right)k^{2}-\mu^{2}\right]k^{2}.\end{equation}
 
 An analysis similar to that done in Sec. III A tells us that the addition of the higher derivative Chern-Simons extension does not improve the non-unitarity of the Lee-Wick-Chern-Simons model. Neither does it cure the non-unitarity of the Lee-Wick system.

\subsection{Higher-derivative gravitational models}
Higher-derivative gravity augmented by a Chern-Simons term is defined by the Lagrangian

\begin{equation}
\mathcal{L}=\sqrt{g}\left(\alpha R+\beta R_{\mu\nu}R^{\mu\nu}+\gamma R^{2}\right)+\frac{\mu}{2}\mathcal{L}_{CS},\label{eq:LagrangianHigherDerCS}\end{equation}
 where \begin{equation}
\mathcal{L}_\mathrm{CS}=\varepsilon^{\mu\nu\rho}\Gamma_{\mu\sigma}^{\lambda}\left(\partial_{\nu}\Gamma_{\rho\lambda}^{\sigma}+\frac{2}{3}\Gamma_{\nu\lambda}^{\kappa}\Gamma_{\rho\kappa}^{\sigma}\right)\end{equation}
is the Chern-Simons term and $\alpha$, $\beta$, $\gamma$, and $\mu$ are arbitrary coefficients.

In the weak field approximation  ($g_{\mu\nu}=\eta_{\mu\nu}+h_{\mu\nu}$), this Lagrangian reduces to

\begin{eqnarray}
\mathcal{L}_{\left(2\right)} &=& \frac{\alpha}{2}\Big(-\frac{1}{2}h^{\mu\nu}\square h_{\mu\nu}+\frac{1}{2}h\square h-h\partial_{\mu}\partial_{\nu}h^{\mu\nu} \nonumber \\ &&+h^{\mu\nu}\partial_{\mu}\partial_{\rho}h_{\ \nu}^{\rho}\Big)
+  \frac{\beta}{4}\Big(h^{\mu\nu}\square^{2}h_{\mu\nu}+h\square^{2}h \nonumber \\ &&- 2h\square\partial_{\mu}\partial_{\nu}h^{\mu\nu} \nonumber\\
 &&- 2h^{\mu\nu}\square\partial_{\mu}\partial_{\rho}h_{\nu}^{\ \rho}+2h^{\mu\nu}\partial_{\mu}\partial_{\nu}\partial_{\rho}\partial_{\sigma}h^{\rho\sigma}\Big)\nonumber \\
&&+  \gamma\Big(h\square^{2}h-2h\square\partial_{\mu}\partial_{\nu}h^{\mu\nu}+h^{\mu\nu}\partial_{\mu}\partial_{\nu}\partial_{\rho}\partial_{\sigma}h^{\rho\sigma}\Big)\nonumber \\
&&+  \frac{\mu}{4}h_{\mu}^{\ \nu}\varepsilon^{\mu\lambda\rho}\partial_{\lambda}\Big(\square h_{\nu\rho}-\partial_{\nu}\partial_{\sigma}h_{\rho}^{\ \sigma}\Big).
\end{eqnarray}

We have now all the ingredients  to compute   the wave operator $\mathcal{O}_{\mu\nu,\rho\sigma}$ and expand it  in the appropriate degree of freedom basis with the aid of the identities collected in the Appendix A. The  resulting  matrices  of the coefficients concerning this expansion are

\begin{widetext}

\begin{equation}
a(0)=\left(\begin{array}{cc}
\left(\left(3\beta+8\gamma\right)k^{2}-\alpha\right)k^{2} & 0\\
0 & 0\end{array}\right),\end{equation}
 \begin{equation}
a\left(2\right)=\left(\begin{array}{cc}
\left(\alpha+\beta k^{2}+\mu\sqrt{k^{2}}\right)k^{2} & 0\\
0 & \left(\alpha+\beta k^{2}-\mu\sqrt{k^{2}}\right)k^{2}\end{array}\right).\end{equation}
\end{widetext}  Due to the gauge symmetry of the model the spin-0 matrix above is evidently non-invertible which is translated  into the usual source conservation constraint on the  gravitational sources $k_{\mu}\mathcal{S}^{\mu\nu}=0$.

On the other hand, the inverse ($A\left(0\right)_{ij}^{-1}$) of the largest nondegenerate matrix extracted from $a\left(0\right)$,
as well as the inverse of $a(2)_{ij}^{-1}$, can be written as
\begin{widetext}

\begin{equation}
A(0)^{-1}=\frac{1}{\left[\left(3\beta+8\gamma\right)k^{2}-\alpha\right]k^{2}},\label{eq:spin 0}\end{equation}
 \begin{equation}
a\left(2\right)^{-1}=\frac{1}{\left[\left(\alpha+\beta k^{2}\right)^{2}-\mu^{2}k^{2}\right]k^{2}}\left(\begin{array}{cc}
\alpha+\beta k^{2}-\mu\sqrt{k^{2}} & 0\\
0 & \alpha+\beta k^{2}+\mu\sqrt{k^{2}}\end{array}\right).\label{eq:spin2}\end{equation}
\end{widetext}

Using the constraints \eqref{eq:ghost/tachyon conditions for massive poles}
on the matrices \eqref{eq:spin 0}-\eqref{eq:spin2}, the following relations for the parameters are obtained

\begin{align}
 & \mbox{Spin-}\mathbf{2}:\ \alpha<0,\ \ \beta>0;\\
 & \mbox{Spin-}\mathbf{0}:\ \alpha>0,\ \ 3\beta+8\gamma>0.\end{align}

Accordingly, for arbitrary values of the parameters the model is  non-unitary.
Nevertheless, there exists a  way of  circumventing  this difficult: all we have to do is to prevent the propagation of
the massive spin- $0$ mode by choosing  $3\beta+8\gamma=0$. Remarkably,
this is precisely the constraint utilized by Bergshoeff, Hohm, and Townsend
(BHT)  in the construction of their model \cite{19}. Another alternative is to inhibit the propagation
of the massive spin-2 mode by setting $\beta=0$. As a consequence,
\begin{equation}
A(0)^{-1}=\frac{1}{8\gamma k^{2}\left(k^{2}-\frac{\alpha}{8\gamma}\right)}.\end{equation}
 If we take into account that  the absence of tachyons and ghosts requires respectively that $\frac{\alpha}{\gamma}>0$ and 
 $\gamma>0$, we come to the  conclusion  that the model 
 $\alpha R+\gamma R^{2}$ is unitary if  $\alpha$
and $\gamma$ are positive.

For the massless poles, one must use the original expression \eqref{eq:propagatorDegenaraciesParityBreaking}
for the propagator in order to compute the residue. The constraints
satisfied by the sources allow us to handle correctly the singularities.
Using such constraints and discarding terms that do not contribute
to the residue, yields

\begin{eqnarray}
\Pi &=&\frac{1}{\alpha k^{2}}i\mathcal{S}^{\ast\mu\nu}  \Big[\frac{1}{2}\left(\eta_{\mu\rho}\eta_{\nu\sigma}+\eta_{\mu\sigma}\eta_{\nu\rho}-\eta_{\mu\nu}\eta_{\rho\sigma}\right) \nonumber \\
 &&- i\mu\varepsilon_{\mu\rho\lambda}\eta_{\nu\sigma}k^{\lambda}\Big]\mathcal{S}^{\rho\sigma}.
\end{eqnarray}

Using a suitable basis for the expansion of the sources in momentum space, we arrive at the conclusion
that this expression vanishes identically, which clearly shows
that the massless mode is non-propagating.

The preceding analysis seems to indicate the existence of two unitary  higher-derivative gravity models in 3D: the BHT and the  $\alpha R+\gamma R^2$ systems. Actually, only the BHT  model can be  really considered a higher-derivative gravity system. Indeed, this model contains fourth-derivatives of the metric, while the pure scalar curvature system is conformally equivalent to Einstein gravity  with a scalar field \cite {Whitt:1984}, which means that despite having fourth derivatives at the metric level the $\alpha R + \gamma R^2$  model is ultimately second-order in its scalar-tensor version.

We point out that the description of gravitational phenomena via the BHT model does not lead to  some really bizarre results as in the usual 3D gravity (lack of a gravity force in the nonrelativistic limit, gravitational deflection independent of the impact parameter, complete absence of gravitational time dilation, no time delay). Actually, in the framework of New Massive Gravity, short-range gravitational forces are exerted on slowly moving particles; besides, the light bending depends on the impact parameter, as it should \cite{41}. And more, both time delay and spectral shift do take place in the context of the alluded model \cite{42}.

Until very recently it was believed that ``New Massive Gravity'' \cite{20,21}  was the  only higher-derivative gravity model in 3D that was simultaneously   perturbatively renormalizable and unitary in flat space \cite {43}.  In a sense, it was expected that   most likely the full model would be  non renormalizable since it only improved the spin-2 projections of the propagator but not the spin-0 projection. We remark, however, that it  was recently shown   that  the general theory of higher-derivative gravity  in 3D is renormalizable, with two notable exceptions  \cite{44,45}: the models in which the coefficients are restricted to the special values $3\beta + 8\gamma=0$ or $\beta=0$. Unfortunately,  those are precisely the systems that are tree-level unitary. Consequently, despite being unitary at the tree level, neither the BHT system nor the $\alpha R + \gamma R^2$ model are renormalizable. Interestingly, this result seems to indicate that the conjecture that unitarity and renormalizability cannot coexist simultaneously in the framework of  one and the same  higher-derivative theory is correct.

It is worth noting that depending on the choice of the parameters in the action concerning higher-derivative gravity in 2+1 dimensions (HDG), one obtains gravity, antigravity or gravitational shielding; in addition, we can analyze the gravitational properties  of HDG using a model somewhat analogous to this one: a plane orthogonal to  a straight cosmic string described by higher-derivative gravity in 3+1 dimensions \cite{46,47}.        


\subsubsection{Higher-derivative-Chern-Simons gravity enlarged by the Ricci-Cotton tensor} 
For  gravity theories  there is  also the possibility of the construction
of a higher derivative Chern-Simons extension, the so-called   Ricci-Cotton term, which is defined by the Lagrangian \begin{equation}
\mathcal{L}_\mathrm{RC}=\lambda\varepsilon^{\mu\nu\rho}R_{\mu\sigma}D_{\nu}R_{\rho}^{\ \sigma}.\end{equation}
 We  remark that models including this term were recently investigated by Bergshoeff, Hohm, and Townsend in their researches on higher derivatives in 3D gravity  and higher-spin gauge theories \cite{Bergshoeff:2010}. 
 
Let us then probe the unitarity of higher-derivative-Chern-Simons gravity augmented via the Ricci-Cotton term. It is curious that this term  only alters the spin-2 sector of this model. The matrix of the coefficients and its inverse are now given by

\begin{widetext}

\begin{equation}
a\left(2\right)=\left(\begin{array}{cc}
\left[\alpha+\beta k^{2}-\left(\mu+\lambda k^{2}\right)\sqrt{k^{2}}\right]k^{2} & 0\\
0 & \left[\alpha+\beta k^{2}+\left(\mu+\lambda k^{2}\right)\sqrt{k^{2}}\right]k^{2}\end{array}\right),\end{equation}

\begin{equation}
a\left(2\right)^{-1}=\frac{1}{\Delta}\left(\begin{array}{cc}
\alpha+\beta k^{2}+\left(\mu+\lambda k^{2}\right)\sqrt{k^{2}} & 0\\
0 & \alpha+\beta k^{2}-\left(\mu+\lambda k^{2}\right)\sqrt{k^{2}}\end{array}\right),\end{equation}
\end{widetext}

\noindent where 
$$\Delta=\left[-\lambda^{2}k^{6}+\left(\beta^{2}-2\mu\lambda\right)k^{4}+\left(2\alpha\beta-\mu^{2}\right)k^{2}+\alpha^{2}\right]k^{2}.$$

A cursory glance at the equation above  is sufficient to convince us that the model at hand can describe at most three massive particles. Proceeding in the same way as we have done in Sec. III A  we conclude   that  the addition of the Ricci-Cotton  term is not a good therapy for curing the nonunitarity of higher-derivative-Chern-Simons gravity.

\section{Concluding Remarks and Comments}\label{sec:Concluding-Remarks}
We have devised an easy procedure for  checking the unitarity of 3D models based on a new class of spin projection operators. The great importance of these operators resides precisely in the fact that they form a complete (and, of course, closed)  set for Lorentz preserving (PV and PC) models. In other words, they obey a completeness relation.  However,  it may happen that a subset of a  given complete set of operators is closed.   A natural and important point to be discussed  in this case is whether or not this    ``incomplete'' set is appropriated  for expanding the propagator. Consider, in this vein, three-dimensional PC gravity models. Now, as we have already commented, the suitable basis for computing the propagator is that whose elements are the well-known Barnes-Rivers operators. Expanding the wave operator in this basis, we obtain in momentum space

\begin{eqnarray}
{\cal{O}}&=&x_1P(1) + x_2P(2) +x_sP(0^{s}) +x_{\omega}P(0^{\omega}) \nonumber  \\   &&+ x_{s \omega}P(0^{s\omega}) +x_{\omega s}P(0^{\omega s}).
\end{eqnarray}

Consequently, the corresponding propagator is given by
 \begin{eqnarray}
\mathcal{O}^{-1}&=& \frac{1}{x_2}P(2) + \frac{1}{x_1}P(1) + \frac{1}{x_s x_ \omega  -x_{s \omega}x_{\omega s}} \nonumber \\ &&\times \Big[ x_\omega P(0^{s}) + x_s P(0^{\omega})  - x_{s \omega} P(0^{s \omega}) \nonumber \\ &&- x_{\omega s}P(0^{\omega s})\Big].
\end{eqnarray}

Now, if $S\equiv \{ P(1), P(2), ... \;,P(0^{\omega s}) \}$, then $S' \equiv \{P(1), P(2), P(0^{s}), P(0^{\omega})\}$ is a subset of $S$ which is closed under the same operation of multiplication as that concerning  $S$; in addition, the elements of $S'$ obey the  relation

\begin{eqnarray}
 P(1)+  P(2) + P(0^{s})+ P(0^{\omega})=\delta,
\end{eqnarray}  

\noindent which is nothing but the decomposition of unity. The relevant question,
nonetheless, is that  (79) is not a completeness relation for the operators at hand. If $S'$ were a complete set we would arrive at the wrong conclusion that it  should necessarily be a basis for performing our computations; as a  consequence, the propagator for the PC gravity models would assume the form

\begin{eqnarray}
\nonumber{\mathcal{O}}^{-1}_{\mathrm{wrong}}=\frac{1}{x_1} P(1)+ \frac{1}{x_2} P(2) +\frac{1}{x_s} P(0^{s})+\frac{1}{x_\omega} P(0^{\omega}).\\
\end{eqnarray}

Comparing (78) and (80) we come to the conclusion that these expressions  coincide only and if only $x_{s \omega} = x_{\omega s} =0$. Nevertheless, these coefficients cannot be zero. Indeed, since the PC gravity models are gauge invariant, we have to add to the Lagrangian of the model a gauge-fixing Lagrangian (${\cal{L}}_\mathrm{gf}$) so that the resulting wave operator can be inverted. Choosing for this purpose, without any loss of generality,    the  de Donder gauge  and taking into account that its linearized version can be written as follows

\begin{eqnarray}   
{\mathcal{L}}_\mathrm{gf} = \frac{1}{2 \lambda}\left(\partial_\mu \gamma^{\mu} \right)^2,
\end{eqnarray}

\noindent where $\gamma^{\mu}= \partial_\nu h^{\mu \nu} - \frac{1}{2} \partial^\mu h$, we promptly obtain in momentum space
 
\begin{eqnarray}
\mathcal{O}_\mathrm{gf} (k)&=& \frac{k^2}{2}\Big[ \frac{1}{2}P(1) + \frac{1}{2}P(0^{s}) +\frac{1}{4} P(0^{\omega}) \nonumber \\ &&- \frac{\sqrt{2}}{4} P(0^{s\omega}) - \frac{\sqrt {2}}{4} P(0^{\omega s}) \Big],
\end{eqnarray}

\noindent which clearly shows that both $x_{s \omega}$ and $x_{\omega s}$ are different from zero. In other words, the operator ${\cal{O}}_\mathrm{wrong} \equiv x_1 P(1) + x_2 P(2) + x_s P(0^{s}) + x_\omega P(0^{\omega})$ is obviously no invertible. Suppose, however, that we argue  that both expressions for the propagator are correct due to the fact that for physical problems in which the propagator (78) is contracted with conserved external currents ($\mathcal{S} {\cal{O}}^{-1}\mathcal{S},  k\mathcal{S}=0$), both operators $P(0^{s\omega})$ and $P(0^{\omega s})$ do not contribute for the final result of the calculations. Again, it is trivial to show that this argument is fallacious. In fact, a straightforward computation leads to the following results

\begin{eqnarray}
\mathcal{S} {\mathcal{O}}^{-1}\mathcal{S}&=& \mathcal{S}\Big[\frac{1}{x_2}P(2) + \frac{x_\omega}{x_s x_\omega - x_{s \omega} x_{\omega s}}P(0^{s})\Big]\mathcal{S}, \nonumber \\ \mathcal{S} {\mathcal{O}}^{-1}_\mathrm{wrong}\mathcal{S} &=& \mathcal{S} \Big[\frac{1}{x_2}P(2) + \frac{1}{x_s}P(0^{s}) \Big]\mathcal{S} \nonumber.
\end{eqnarray}

 In summary,  only  operators that form a complete set can be used to attaining the propagator. In other words,  the term ``closure'' cannot be used as a  synonym for ``completeness''.  The fact the our three-dimensional set of operators is  indeed complete   thus  guarantees that the physical results obtained through  them  can be trusted.

Another point that deserves to be discussed is whether  the nonunitary disease that affects some three-dimensional models could be cured by the addition of a Chern-Simons  term  to the system; of course, we are not excluding from our considerations the possibility of enlarging the model via a higher derivative Chern-Simons extension or even through the simultaneous addition of Chern-Simons and higher derivative Chern-Simons terms. 
 For the sake of brevity, we restrict our analysis to three-dimensional gravitational models.  Everything started when it was found out that the solution  to the triviality problem of   general relativity in (2+1)D could be cured by simply adding a topological Chern-Simons term to the system. The resulting model describes a  non-trivial gravity theory with a propagating, massive, spin-2 mode \cite{7}. Later on  it was considered another way out of the triviality problem of 3D gravity: the addition of higher-derivative terms to the system \cite{Accioly:2001}; unfortunately the resulting  models  are nonunitary \cite{Accioly:2002}. On the other hand, it was  claimed  that the addition of a Chern-Simons term to the previous model would cure its non-unitarity \cite{Pinheiro:1997}. This was proved afterwards to be incorrect \cite{Accioly:2000}. After this digression, let us respond the question we have raised above. As we have seen in Section III, nonunitary higher-derivative electromagnetic (gravitational) models do not become unitary systems by simply augmenting them  through Chern-Simons terms. Neither do they become unitary by enlarging them via a higher derivative Chern-Simons extension.  It is amazing, nonetheless, that are some examples in the literature of unitary systems whose unitarity is spoiled by the addition of Chern-Simons terms \cite{Accioly:2004, Accioly:2005}. Therefore, in some cases, the coexistence between the topological term and higher-derivative theories is conflicting. Consequently, the addition of a Chern-Simons term to a given model should be based on transparent physical results. This was precisely the most important criterion we have adopted for choosing the models discussed in the text.
 
 To conclude we would like  to point out that the results of this work are actually relevant to graphene.  In fact, recently quantum field theory methods have been applied to analyze the properties of this interesting system.  As a consequence, the Dirac model from the tight binding model was derived and calculations of the polarization operator (conductivity) were described.  Subsequently, this polarization operator  was used to describe the Quantum Hall Effect, light absorption by graphene, the Faraday effect, and the Casimir interaction \cite{Fialkovsky:2012}. There are also interesting studies of the graphene with emphasis on Chern-Simons terms \cite{Santangelo:2008,Beneventano:2009}.   The use of our three-dimensional operators will certainly benefit the  computations involving   the polarization operator, as well as those   related to  models augmented by  Chern-Simons terms.

\section*{Acknowledgements}

The authors are very grateful to  CNPq (Brazilian agency) for
financial support.

\paragraph*{Note added.}

S.  Kruglov, whom we thank,  informed us about an interesting  class of 
spin operators for  Maxwell-Chern-Simons topologically massive
gauge fields in 3D that he   recently  found out \cite{Kruglov:2010at}.

\appendix

\section{Projection Operators and Tensor Relations\label{sec:ProjectionOperatorRelationAppendix}}

In this appendix, we gather the spin operators constructed
in Sec. \ref{sec:A-new-set} and some useful identities satisfied
by them.

\subsection{Vector field operators: $A-A$}

\paragraph{Spin-0 Sector}
\begin{itemize}
\item $P^{AA}\left(0\right)_{\mu\nu}=\omega_{\mu\nu}$ 
\end{itemize}

\paragraph{Spin-1 Sector}
\begin{itemize}
\item $P_{11}^{AA}\left(+1\right)_{\mu\nu}=\rho_{\mu\nu}$ 
\item $P_{22}^{AA}\left(-1\right)_{\mu\nu}=\sigma_{\mu\nu}$ 
\end{itemize}

\paragraph{Identities Among the Operators}
\begin{itemize}
\item $P^{AA}\left(1\right)_{\mu\nu}=\theta_{\mu\nu}=P_{11}^{AA}\left(+1\right)_{\mu\nu}+P_{22}^{AA}\left(-1\right)_{\mu\nu}$ 
\end{itemize}

\paragraph{Tensorial Identities }

\begin{eqnarray}
 & \eta_{\mu\nu} & =P^{AA}\left(0\right)_{\mu\nu}+P^{AA}\left(1\right)_{\mu\nu} \nonumber\\
 & k_{\mu}k_{\nu} & =k^{2}P^{AA}\left(0\right)_{\mu\nu} \nonumber \\
 & \varepsilon_{\mu\nu\rho}k^{\rho} & =i\sqrt{k^{2}}\left(P_{11}^{AA}\left(+1\right)_{\mu\nu}-P_{22}^{AA}\left(-1\right)_{\mu\nu}\right) \nonumber\end{eqnarray}

\begin{widetext}

\subsection{Rank-2 Symmetric Field Operators: $h-h$ }

\paragraph{Spin-0 sector}
\begin{itemize}
\item $P_{11}^{hh}\left(0^{s}\right)_{\mu\nu;\rho\sigma}=\frac{1}{2}\theta_{\mu\nu}\theta_{\rho\sigma},$ 
\item $P_{22}^{hh}\left(0^{\omega}\right)_{\mu\nu;\rho\sigma}=\omega_{\mu\nu}\omega_{\rho\sigma},$ 
\item $P_{12}^{hh}\left(0^{s\omega}\right)_{\mu\nu;\rho\sigma}=\frac{1}{\sqrt{2}}\theta_{\mu\nu}\omega_{\rho\sigma},$ 
\item $P_{21}^{hh}\left(0^{\omega s}\right)_{\mu\nu;\rho\sigma}=\frac{1}{\sqrt{2}}\omega_{\mu\nu}\theta_{\rho\sigma}.$ 
\end{itemize}

\paragraph{Spin-1 Sector}
\begin{itemize}
\item $P_{11}^{hh}\left(+1\right)_{\mu\nu;\rho\sigma}=\frac{1}{2}\left(\rho_{\mu\rho}\omega_{\nu\sigma}+\rho_{\nu\rho}\omega_{\mu\sigma}+\rho_{\mu\sigma}\omega_{\nu\rho}+\rho_{\nu\sigma}\omega_{\mu\rho}\right),$ 
\item $P_{22}^{hh}\left(-1\right)_{\mu\nu;\rho\sigma}=\frac{1}{2}\left(\sigma_{\mu\rho}\omega_{\nu\sigma}+\sigma_{\nu\rho}\omega_{\mu\sigma}+\sigma_{\mu\sigma}\omega_{\nu\rho}+\sigma_{\nu\sigma}\omega_{\mu\rho}\right),$ 
\item $P_{12}^{hh}\left(\pm1\right)_{\mu\nu;\rho\sigma}=\frac{1}{2}\varepsilon_{\tau\eta\kappa}\left(\rho_{\mu}^{\tau}\sigma_{\rho}^{\eta}\omega_{\nu\sigma}+\rho_{\nu}^{\tau}\sigma_{\rho}^{\eta}\omega_{\mu\sigma}+\rho_{\mu}^{\tau}\sigma_{\sigma}^{\eta}\omega_{\nu\rho}+\rho_{\nu}^{\tau}\sigma_{\sigma}^{\eta}\omega_{\mu\rho}\right)\frac{k^{\kappa}}{\sqrt{k^{2}}},$ 
\item $P_{21}^{hh}\left(\mp1\right)_{\mu\nu;\rho\sigma}=\frac{1}{2}\varepsilon_{\tau\eta\kappa}\left(\sigma_{\mu}^{\eta}\rho_{\rho}^{\tau}\omega_{\sigma\nu}+\sigma_{\mu}^{\eta}\rho_{\sigma}^{\tau}\omega_{\rho\nu}+\sigma_{\nu}^{\eta}\rho_{\rho}^{\tau}\omega_{\sigma\mu}+\sigma_{\nu}^{\eta}\rho_{\sigma}^{\tau}\omega_{\rho\mu}\right)\frac{k^{\kappa}}{\sqrt{k^{2}}}.$ 
\end{itemize}

\paragraph{Spin-2 Sector}
\begin{itemize}
\item $P_{11}^{hh}\left(+2\right)_{\mu\nu;\rho\sigma}=\rho_{\mu\rho}\rho_{\nu\sigma},$ 
\item $P_{22}^{hh}\left(-2\right)_{\mu\nu;\rho\sigma}=\sigma_{\mu\rho}\sigma_{\nu\sigma}.$ 
\end{itemize}

\paragraph{Identities Among the Operators}
\begin{itemize}
\item $P^{hh}\left(1\right)_{\mu\nu;\rho\sigma}=\frac{1}{2}\left(\theta_{\mu\rho}\omega_{\nu\sigma}+\theta_{\nu\rho}\omega_{\mu\sigma}+\theta_{\mu\sigma}\omega_{\nu\rho}+\theta_{\nu\sigma}\omega_{\mu\rho}\right)=P_{11}^{hh}\left(+1\right)+P_{22}^{hh}\left(-1\right)$ 
\item $P^{hh}\left(2\right)_{\mu\nu;\rho\sigma}=\frac{1}{2}\left(\theta_{\mu\rho}\theta_{\nu\sigma}+\theta_{\mu\sigma}\theta_{\nu\rho}-\theta_{\mu\nu}\theta_{\rho\sigma}\right)=P_{11}^{hh}\left(+2\right)+P_{22}^{hh}\left(-2\right)$ 
\end{itemize}

\paragraph{Tensorial Identities }

\begin{align}
 & \delta_{\mu\nu,\rho\sigma}=\frac{1}{2}\left(\eta_{\mu\rho}\eta_{\nu\sigma}+\eta_{\mu\sigma}\eta_{\nu\rho}\right)=P^{hh}\left(2\right)+P^{hh}\left(1\right)+P_{11}^{hh}\left(0\right)+P_{22}^{hh}\left(0\right) \nonumber\\
 & \eta_{\mu\nu}\eta_{\rho\sigma}=2P_{11}^{hh}\left(0^{s}\right)+\sqrt{2}P_{12}^{hh}\left(0\right)+\sqrt{2}P_{21}^{hh}\left(0\right)+P_{22}^{hh}\left(0\right) \nonumber \\
 & k_{\mu}k_{\nu}\eta_{\rho\sigma}+k_{\rho}k_{\sigma}\eta_{\mu\nu}=\sqrt{2}k^{2}\left(P_{12}^{hh}\left(0\right)+P_{21}^{hh}\left(0\right)\right)+2k^{2}P_{22}^{hh}\left(0\right) \nonumber\\
 & k_{\mu}k_{\rho}\eta_{\nu\sigma}+k_{\mu}k_{\sigma}\eta_{\nu\rho}+k_{\nu}k_{\rho}\eta_{\mu\sigma}+k_{\nu}k_{\sigma}\eta_{\mu\rho}=2k^{2}P^{hh}\left(1\right)+4k^{2}P_{22}^{hh}\left(0\right) \nonumber\\
 & k_{\mu}k_{\nu}k_{\rho}k_{\sigma}=k^{4}P_{22}^{hh}\left(0\right) \nonumber\\
 & \left(\varepsilon_{\kappa\rho\mu}\eta_{\nu\sigma}+\varepsilon_{\kappa\rho\nu}\eta_{\mu\sigma}+\varepsilon_{\kappa\sigma\mu}\eta_{\nu\rho}+\varepsilon_{\kappa\sigma\nu}\eta_{\mu\rho}\right)k^{\kappa}= \nonumber\\
 & 2\sqrt{k^{2}}\left(P_{11}^{hh}\left(+2\right)-P_{22}^{hh}\left(-2\right)-P_{12}^{hh}\left(\pm1\right)+P_{21}^{hh}\left(\mp1\right)\right)\nonumber \\
 & \left(\varepsilon_{\kappa\rho\mu}k_{\nu}k_{\sigma}+\varepsilon_{\kappa\rho\nu}k_{\mu}k_{\sigma}+\varepsilon_{\kappa\sigma\mu}k_{\nu}k_{\rho}+\varepsilon_{\kappa\sigma\nu}k_{\mu}k_{\rho}\right)k^{\kappa}=2k^{2}\sqrt{k^{2}}\left(-P_{12}^{hh}\left(\pm1\right)+P_{21}^{hh}\left(\mp1\right)\right) \nonumber\end{align}

Here the $\mu\nu;\rho\sigma$ indices of the operators $P_{ij}^{hh}\left(J\right)_{\mu\nu;\rho\sigma}$ were omitted.\end{widetext}









\begin{thebibliography}{50}
\bibitem{2}{Graphene is a flat, two-dimensional system, consisting of a mono-layer of carbon atoms in a honeycomb array.}
\bibitem{34}{These models play an important role as far as the mechanism of topological
mass generation of photons (gravitons) is concerned.}
\bibitem{35}{This class of operators spans the vector space constituted by the
linear operators that act upon the vector representation of the Lorentz
group.}
\end{thebibliography}
\end{document}